\documentclass[pra,twocolumn, 10pt, notitlepage, nofootinbib]{revtex4-2}
\usepackage{amsmath, amsfonts, amssymb, array, amsthm, bm}
\usepackage{mathrsfs}
\usepackage{physics}
\usepackage{graphicx}
\usepackage{color}

\begin{document}

\title{Dynamical Maps for Accelerating Detectors}

\author{Shalin Jose} 
\email{shaliniiser16@iisertvm.ac.in}
\affiliation{Indian Institute of Science Education and Research Thiruvananthapuram, Vithura, Kerala, 695551, India} 

\author{Anil Shaji}
\affiliation{Indian Institute of Science Education and Research Thiruvananthapuram, Vithura, Kerala, 695551, India}

\begin{abstract}
We study the open quantum dynamics of a two-level particle detector that starts accelerating through Minkowski vacuum weakly coupled to a massless scalar field. We consider a detector with non-zero size and study its time evolution for the case where it is initially in inertial motion and subsequently a constant acceleration is switched on for a finite time. We study the dynamical maps that describe the evolution of such a system and show that the dynamics is not completely positive (NCP). The inertial motion prior to the acceleration can entangle the detector and field leading to the NCP dynamics. We examine the nature of the open dynamics during the accelerated phase as a function of the duration of prior inertial motion and the magnitude of the acceleration. 
\end{abstract}

\maketitle

\section{Introduction \label{intro}}

Mathematical description of the dynamics of open quantum systems is typically in terms of completely positive dynamical maps~\cite{ECGPhysRev.121.920,choi75,Choi1972,BreuerBook,kraus83b,Rivas_2012,nielsen_chuang_2010,Erling1963}. The seminal paper on dynamical maps however does not require them to be completely positive and instead uses the physically motivated restriction that the action of such maps on quantum states in its domain must be positivity preserving~\cite{ECGPhysRev.121.920}. The stronger requirement of complete positivity was introduced subsequently and since then there has been quite a bit of discussion on whether complete positivity is really needed or not~\cite{ShajiInitialEntangled,pechukas94a,shaji_whos_2005,stormer_completely_2010,laine_witness_2010,wood_non-completely_2009,Rodr_guez_Rosario_2008}. Apart from experimental observations of dynamics that are not completely positive, an interesting case in support of abandoning the requirement of complete positivity arises when changes in state induced by relativistic coordinate transformations turn out to be not completely positive~\cite{jordan05a}. In such cases, the lack of complete positivity is to be treated as a `kinematic' effect that cannot be avoided unless very restrictive initial conditions are assumed. In this paper we examine a similar scenario that appears in the context of the Unruh effect. 

A particle detector sensitive to excitations of a massless scalar field, accelerating through vacuum, will see a thermal state of the field in its accelerating frame of reference. If the detector is undergoing uniform acceleration it will reach thermal equilibrium with a Planck distribution of (Rindler) excitations of the field which we refer to as `photons' for brevity. The equilibrium temperature felt by the detector is directly proportional to its acceleration, $a$, and is given by the relation $T = \hbar a /2 \pi c k_{B}$. This prominent result first described in~\cite{Unruh} is referred to as the Unruh effect and has become one of the key theoretical results to come out of introducing quantum field theoretic (QFT) considerations in quantum information theory. Unruh effect introduces the idea of the vacuum of the quantum field as observer dependent for non-inertial frames. The detector that responds to quanta of the field is typically modeled by a point-like, two-level system which is referred to as the Unruh-DeWitt Detector~\cite{Unruh, Dewitt}. Since its discovery, there has been growing interest in the Unruh effect \cite{ApplicationsUnruh} and the challenge of demonstrating it experimentally. This has been a difficult task since even low detectable temperatures correspond to very high accelerations. Some of the suggested experimental proposals that could potentially lead to detection of this effect can be found in~\cite{exp1, berry, Pena2014, coldatoms, circle}. However, none of these proposals have been implemented in the lab so far. In~\cite{HawkingOQS}, the Hawking radiation has been studied as an open quantum system and its similarities to the Unruh effect provide another motivation to study accelerating systems.

The Unruh-Dewitt detector was originally introduced as a construct that allows one to approach the question of how inertial or accelerated observers see the quanta of a field by asking whether the particle detector will `click' or not. The detector is also sensitive to the vacuum fluctuations of the scalar field. The contribution to the excitation probability of an inertial detector, initially in its ground state, from the vacuum fluctuations averages out to zero over long timescales~\cite{birrell_davies_1982}. However, for short time scales, the contribution is not negligible. The inertial and accelerated excitation probabilities for a point-like detector for finite times were obtained in \cite{Svaiter}. In other words, a detector that has been initialised in the sense that it has been prepared in a specific state and its interaction with the field has been switched on, but is otherwise stationary (inertial), has an `unavoidable' and non-trivial interaction with the field even if the field is in the vacuum state. Due to this interaction, quantum correlations can develop between the two. Subsequent acceleration of the detector induces detector dynamics that need not be completely positive because of these initial quantum correlations.

 Since the focus of our discussion is the detector itself, we would like to treat it as an experimentally accessible quantum system with finite size. A detector with finite size was first considered in \cite{Tagaki} and later by Schlicht in \cite{Schlicht_2004} which is the approach we follow. In a more recent work \cite{Sudhir_2021}, the effect of finite mass of the detector has also been studied bringing the idealised Unruh-Dewitt detector closer to a real physical system. In \cite{inertialspherical}, the inertial excitation probability calculated in \cite{Svaiter} is re-derived with adiabatic switch-on of the coupling in the infinite past. In \cite{B.L.Hu} non-perturbative analysis is used to show that even in the case of inertial motion there is entanglement between the detector and the field which is an important point in our study. Previously, accelerating detectors have been analysed from the perspective of open quantum dynamics in \cite{Benatti,timerange}. Most of the existing literature on the Unruh Effect is primarily on eternally accelerating detectors. Some recent papers have explored various other trajectories \cite{Louko_2006}, including time-dependent ones \cite{timedepacc}. 

In the following we describe the dynamics of the accelerated, realistic detector in terms of the dynamical map induced on its internal state due to its motion in the presence of the field. After initialisation of the two-level detector, its  subsequent motion may be inertial or non-inertial. Specifically we are interested in a trajectory similar to the one used in \cite{Sabrina2020} wherein the detector, after initialisation, is at rest for a certain period of time before it undergoes motion in one dimension with constant acceleration. However, unlike the scenario considered in~\cite{Sabrina2020} where detector initialisation happens at $t=-\infty$ and the detector is always on, we choose to initialise it at a finite time prior to the start of the accelerated motion. Even though the initial stationary (inertial) period may be uninteresting from the point of view of physical motion of the detector, as mentioned earlier, it can still affect the nature of dynamics of the quantum state when it undergoes accelerated motion subsequently because of the influence of the vacuum fluctuations of the field on the stationary (inertial) detector. 

Quantum master equations have been used in \cite{Sabrina2020, Dimitris} to study the Markovian and Non-Markovian behaviour of accelerating detectors. Particularly, in \cite{Sabrina2020} the complete positivity conditions \cite{SabrinaCP} are observed to be broken without a clear explanation as to why it is broken. The discussion in \cite{Sabrina2020} considers the decay rates appearing in the master equation describing detector evolution. We study the finite time dynamics \cite{Sriramkumar1996} of a similar system with the aim of understanding the not completely positive nature of the evolution. We show how this is an expected result if we take into account the inertial part of the trajectory. For this, we split the evolution of the detector into two parts consisting of inertial and accelerated motion after which we determine the dynamical map describing only the accelerated evolution. We obtain analytic expressions for the state of the detector at various points of time that are of interest. This allows us to determine linear dynamical maps describing each stage of the evolution. The duration of time \cite{timerange} for which our results apply are limited to the range where the commonly used perturbative expansion \cite{birrell_davies_1982} is valid. The perturbative approach means that the dynamical maps are not necessarily trace preserving but in the ranges of time we consider, the trace corrections to the density matrix are found to be negligible or even equal to zero. We use natural units with $c = k_{B} = \hbar = 1$ throughout the Paper. 

This Paper is organised as follows: In Sec.~\ref{Methods} we summarise the key results from~\cite{Svaiter} on the two-level detector system and introduce into this discussion the smeared field operator from~\cite{Schlicht_2004} describing a finite size detector. In the next section we obtain the reduced density matrix for a detector which is either inertial or is uniformly accelerating for a finite time after switch-on. In both cases, the initial state of the detector and field is a product state. In Sec.~\ref{InAc} we calculate the reduced density matrix for a detector that was in inertial motion for a finite time before a constant acceleration is switched on at $\tau =0$. The induced dynamical maps are discussed in Sec.~\ref{dynmap} in terms of the dynamical matrix $B$ for the cases presented in Sections \ref{Accelerated Detector} and \ref{InAc}. We summarise our results and also discuss the limitations of our approach in Sec.~\ref{conclusion}. 

\section{The transition probability}\label{Methods}

The Unruh-Dewitt detector, labelled as $S$, is a qubit with two energy levels $\omega_g$ and $\omega_e$ respectively. The corresponding energy eigenstates are labelled as $|0\rangle$ and $|1\rangle$. The scalar field that forms the environment of the detector is labelled as $E$ and its modes with energy $\omega_k$ are created and annihilated by operators $a_k$ and $a_k^\dagger$ respectively. We closely follow the discussion in~\cite{Svaiter} and consider, at first, a point-like detector and  with the detector-field system described by the Hamiltonian,
\begin{equation}
H = H_{S} + H_{E} + H_{\rm int},
\end{equation} 
where $H_{S}$, $H_{E}$ and $H_{\rm int}$ are the Hamiltonians for the free evolution of the detector, evolution of the free scalar field and interaction between the two respectively. We have
\[ H_{S} =  \omega_{e}|1\rangle \langle 1| + \omega_{g}|0\rangle \langle 0 |, \;\;   {\rm and}  \;\;   H_{E} = \int \frac{d^{3}k}{(2\pi)^{3}} \omega_{k} a^{\dagger}_{k}a_{k}. \]
We describe the motion of the detector from the vantage point of an observer co-moving with it and the motion is parametrised by the proper time $\tau$ of the detector that this observer sees. The world-line of the detector is given by $\underline{x}_\tau = (t, {\bf x})$. Since we will be considering motion in one spatial dimension, the world-line can be written as $\underline{x}_\tau = (t, x)$. The detector is considered to be `active' or `on' when its state is in the qubit subspace spanned by $|0\rangle$ and $|1\rangle$. When active, the detector couples to the scalar field $\phi(\underline{x}_\tau)$ along its world line through the interaction Hamiltonian,
\begin{equation}
H_{\rm int}(\tau) = \hat{m} \phi(\underline{x}_\tau),
\end{equation}
where the detector-field interaction is mediated by the monopole moment $\hat{m}$ of the detector defined as 
\begin{equation}
	\hat{m} = m |1\rangle \langle 0| +  m^*|0\rangle \langle 1|. 
\end{equation}
The coupling strength $|m|$ between the detector and the field is assumed to be weak in all the discussions that follow. 

We work in the interaction picture with 
\begin{equation}
\label{Hinteraction}
H_{I}(\tau) = e^{i H_{0} \tau} H_{\rm int} e^{-i H_{0} \tau} =  \hat{m}(\tau) \phi_I(\underline{x}_\tau),
\end{equation}
where $H_{0} = H_{S} + H_{E}$ and
\begin{equation}
	\hat{m}(\tau) = e^{i \omega \tau} m |1\rangle \langle 0| +  e^{-i \omega \tau} m^*|0\rangle \langle 1|, \quad \omega \equiv \omega_e - \omega_g.
\end{equation}
Only the interaction-picture field operator appears in our discussions and so we simplify the notation and denote $\phi_I(\underline{x}_\tau)$ as $\phi(\tau)$ in the following. Since the detector-field coupling is weak, the unitary time evolution operator describing their joint evolution, between $\tau_i$ and $\tau_f$, in the interaction picture can be approximated by truncating the corresponding Dyson series expansion as, 
\begin{eqnarray}
\label{unitary1}
U(\tau_{f},\tau_{i}) & \approx &  1 - i \int_{\tau_{i}}^{\tau_{f}} d\tau'H_{I}(\tau') \nonumber \\
&& \quad - 
\frac{1}{2} {\mathcal T} \int_{\tau_{i}}^{\tau_{f}} \!\!\!\int_{\tau_{i}}^{\tau_{f}} d\tau' \,  d\tau'' H_{I}(\tau') H_{I}(\tau''), \quad
\end{eqnarray}
where ${\mathcal T}$ stands for the time-ordering operator. 

As a first step, we compute the probability that the detector undergoes a transition from $|0\rangle$ to $|1\rangle$ due its interaction with the Poincare-invariant, Minkowski vacuum state, $|{\bf 0}_M\rangle$, of the field. We have,
\begin{eqnarray}
	\label{psif0}
	|\Psi^0_f\rangle & = & U(\tau_{f},\tau_{i})|0\rangle|{\bf 0}_M\rangle \nonumber \\ 
	& = & |0\rangle \bigg[\openone -\frac{|m|^2}{2}  {\mathcal T}\hat{A}^-(\tau_i, \tau_f) \bigg] |{\bf 0}_M\rangle \nonumber  \\
	&& \qquad \qquad \qquad  -\, im |1\rangle \hat{B}(\tau_i, \tau_f)|{\bf 0}_M\rangle 
\end{eqnarray}
where
\begin{eqnarray}
\label{ABdef}
 \hat{ A}^\pm (\tau_i, \tau_f) & = &   \int_{\tau_{i}}^{\tau_{f}} \!\!\!\int_{\tau_{i}}^{\tau_{f}} d\tau' \,  d\tau'' e^{\pm i\omega(\tau'-\tau'')}  \phi(\tau') \phi(\tau''), \nonumber \\
	\hat{B}(\tau_i, \tau_f) & = & \int_{\tau_{i}}^{\tau_{f}}  d\tau' e^{i\omega \tau'} \phi(\tau')
\end{eqnarray}
We can now obtain the reduced density matrix, $\rho^S$, of the detector as 
\begin{equation}
	\rho^S = {\rm tr}_E\big[ |\Psi_f^0\rangle  \langle \Psi_f^0| \big].
\end{equation}
The transition probability for the detector initialised in state $|0\rangle$ to state $|1\rangle$ is given by 
\begin{eqnarray}
	\label{transprob1}
	\rho^S_{11} & = & |m|^2 \langle {\bf 0}_M|\hat{B}^\dagger(\tau_i, \tau_f)  \hat{B}(\tau_i, \tau_f) |{\bf 0}_M\rangle \nonumber \\
	& = & |m|^2 \langle {\bf 0}_M|\hat{A}^-(\tau_i, \tau_f)   |{\bf 0}_M\rangle \nonumber \\
	& = & |m|^2 \int_{\tau_i}^{\tau_f} \!\!\!\!\int_{\tau_i}^{\tau_f} \!\!\! d\tau' \, d\tau'' \, e^{-i\omega (\tau' - \tau'')} {\mathcal W}_M(\tau',\tau''),  \quad 
\end{eqnarray}
where the two-point correlation (Wightman) function is defined as,
\begin{equation}
	{\mathcal W}_M(\tau',\tau'') = \langle {\bf 0}_M | \phi(\tau') \phi(\tau'') | {\bf 0}_M \rangle.	
\end{equation}
Writing massless scalar field operator as, 
\begin{equation}
\phi(\underline{x}) = \frac{1}{(2\pi)^{3}}\int \frac{d^{3}k}{\sqrt{2 \omega_{k}}}\big(a_{k} e^{-i\underline{k}\,\underline{x}} + a^{\dagger} e^{i\underline{k}\,\underline{x}}\big),
\end{equation}
with $\underline{k} \, \underline{x} = -\omega_{k} t + {\mathbf k} \cdot {\mathbf x}$, choosing a metric signature $(-1,1,1,1)$, and using $a_k|0_M\rangle = \langle 0_M|a_k^\dagger=0$ we obtain, 
\begin{equation}
	\label{wightman1}
    {\mathcal W}_M(\tau',\tau'') = \frac{1}{(2\pi)^{3}}\int \frac{d^{3}k}{2\omega_{k}} e^{-i\underline{k}(\underline{x}_{\tau^\prime} - \underline{x}_{\tau^{\prime \prime}})}
\end{equation}
The integral in the equation above does not converge and a suitable regularisation has to be introduced. 

\subsection{Realistic detectors \label{realistic}}
In \cite{Schlicht_2004} a physical understanding of the infinity appearing in the Wightman function is proposed as stemming from the point-like nature of the detector and its interaction with the field. Instead of using one of the standard methods of regularising the integral in Eq.~\eqref{wightman1} like moving the pole on the real line to the complex plane, a natural approach in this case would be to consider a finite spatial size for the detector. A detector with finite, rigid size interacts with the field not at a point but over an extended region of space. The same effect can be reproduced by considering a `smeared' field operator at each point in space instead. Such a smeared field operator was introduced in \cite{Schlicht_2004,Louko_2006} as,
\begin{equation}
    \Phi(\tau) = \int d^{3}\xi f_{\epsilon}(\bm{\xi}) \phi(x_{\tau, \bm{\xi}})
\end{equation}
Where $\bm{\xi}$ denote local Fermi-Walker coordinates \cite{fermicord} associated with the world-line of the detector. The profile function $f_{\epsilon}(\bm{\xi})$ determines the shape of the detector and its size is characterised by $\epsilon$. A convenient choice for $f_{\epsilon}(\bm{\xi})$ that is used in \cite{Schlicht_2004}, corresponding to a rigid, non-rotating detector is,
\begin{equation}
	\label{profile1}
    f_{\epsilon}(\bm{\xi}) = \frac{1}{\pi^{2}} \frac{\epsilon}{(|\bm{\xi}|^{2} + \epsilon^2)^{2}}
\end{equation}
In the limit of $\epsilon \rightarrow 0$ this choice of profile function becomes the delta function and it reduces to the case of a point detector. The smeared field operator corresponding to the choice of profile function in Eq.~\eqref{profile1} can be written as, 
 \begin{equation}
 \label{smearedfield}
	\Phi(\tau) = \frac{1}{(2\pi)^{3}}\int \frac{d^{3}k}{\sqrt{2 \omega_{k}}}\big[a_{k} g(\underline{k},\tau) + a_k^{\dagger} g^{*}(\underline{k},\tau)\big],
\end{equation}	
where
\begin{equation}
\label{gfunction}
    g(\underline{k},\tau) = \int d^{3}\xi f_{\epsilon}(\bm{\xi}) e^{i\underline{k}\,\underline{x}_{\tau,\bm{\xi}}} =  e^{i \underline{k} \, \underline{x}_{\tau}} e^{\epsilon \underline{k}\, \underline{u}_{\tau}} 
\end{equation}
with $\underline{x}_{\tau}$ and $\underline{u}_{\tau}$ being the position and velocity four vector of the detector respectively. The creation and annihilation operators in~\eqref{smearedfield} obey the standard commutation relations,  $[a^{\phantom{\dagger}}_{k}, a_{k'}^{\dagger}] = (2 \pi)^{3} \delta(k-k')$.  The smeared field operator leads to the regularised Wightman function (See Appendix \ref{AppA} for details), 
\begin{equation}
	\label{wightman2}
	{\mathcal W}_M(\tau',\tau'') = \frac{1/4\pi^{2}}{\big[\underline{x}_{\tau^{\prime }} - \underline{x}_{\tau^{\prime \prime}} - i \epsilon (\underline{u}_{\tau^{\prime }} + \underline{u}_{\tau^{\prime \prime}})\big]^{2}}
\end{equation}
We see that when the detector has a stationary world line with $\mathbf{u} = 0$, then the regularised Wightman function reduces to
\begin{equation}
\label{wightman3}
	{\mathcal W}^{0}_M(\tau',\tau'') = \frac{-1/4\pi^{2}}{(\tau^{\prime} - \tau^{\prime \prime} -2i\epsilon)^2},
\end{equation}
which is also the expression one would obtain through the standard regularisation procedure of moving the pole of the expression on the right hand side of Eq.~\eqref{wightman1} to the complex plane by a distance $2\epsilon$. 
  
\section{Inertial and accelerated motion}\label{Accelerated Detector}

We continue the development in Eq.~(\ref{psif0}) and compute the action of the unitary from Eq.~(\ref{unitary1}) on the initial state $|1\rangle|{\mathbf 0}_M\rangle$ as 
\begin{eqnarray}
	\label{psif1}
	|\Psi^1_f\rangle & = & U(\tau_{f},\tau_{i})|1\rangle|{\mathbf 0}_M\rangle \nonumber \\ 
	& = & |1\rangle \bigg[\openone -\frac{|m|^2}{2}  {\mathcal T}\hat{A}^+(\tau_i, \tau_f) \bigg] |{\bf 0}_M\rangle \nonumber  \\
	&& \qquad \qquad \qquad  -\, im^* |0\rangle \hat{B}^\dagger(\tau_i, \tau_f)|{\mathbf 0}_M\rangle. 
\end{eqnarray}
Following the discussion in  Sec.~\ref{realistic}, we are assuming that the field operators $\phi(\tau)$ appearing in the definitions of operators $\hat{A}^\pm$ and $\hat{B}$ given in Eq.~(\ref{ABdef}) have been replaced with the smeared field operators $\Phi(\tau)$. 
With these replacements we now compute the evolution of the detector starting at time $\tau_i$ in a generic initial state of the form $|\psi_i\rangle |{\mathbf 0}_M\rangle$, where
\begin{equation}
\label{rhoinitial}
|\psi_{i}\rangle = \cos \frac{\theta}{2}  |0\rangle + \sin \frac{\theta}{2}  e^{i \varphi}|1\rangle.
\end{equation}
The state of the detector at time $\tau_f$ is given by
\begin{equation}
\label{rhofinal2}	
\rho^S_{0}(\tau_f) = {\rm Tr} \big[U(\tau_{f},\tau_{i})|\psi_i\rangle |{\mathbf 0}_M\rangle \langle {\mathbf 0}_M|\langle \psi_i| U^\dagger(\tau_{f},\tau_{i}) \big], 
\end{equation}
where the subscript `0' indicates that the detector was subject to inertial motion. Using Eqs.~(\ref{psif0}), (\ref{psif1}) and (\ref{rhofinal2}), we find
\begin{eqnarray}
	\label{rhofinal0}
	\rho^S_{0}(\tau_f) & = & \Big[ \cos^2\frac{\theta}{2} \big( 1 - |m|^2 \Re[{\mathcal T} {\mathcal Y}_{00}^{-+}] \big) \nonumber \\
	&& \qquad \qquad  + \, |m|^2 \sin^2\frac{\theta}{2} {\mathcal Y}_{00}^{+-} \Big] |0\rangle \langle 0| \nonumber \\
	&& + \Big[ \sin^2\frac{\theta}{2} \big( 1 - |m|^2 \Re[{\mathcal T} {\mathcal Y}_{00}^{+-}] \big) \nonumber \\
	&& \qquad \qquad  + \, |m|^2 \cos^2\frac{\theta}{2} {\mathcal Y}_{00}^{-+} \Big] |1\rangle \langle 1| \nonumber \\
	&& + \frac{1}{2} \sin \theta \bigg[  e^{-i\varphi} \bigg\{ \! 1 \! - \!\frac{|m|^2}{2} [{\mathcal T} {\mathcal Y}_{00}^{-+}+({\mathcal T} {\mathcal Y}_{00}^{+-})^*] \! \bigg\} \nonumber \\
	&& \qquad \qquad  \qquad  + e^{i\varphi} (m^*)^2 {\mathcal Y}_{00}^{--}\bigg] |0\rangle \langle 1| \nonumber \\
	&& + \frac{1}{2} \sin \theta \bigg[  e^{i\varphi} \bigg\{ 1 -\frac{|m|^2}{2} [({\mathcal T} {\mathcal Y}_{00}^{-+})^*+{\mathcal T} {\mathcal Y}_{00}^{+-}] \bigg\} \nonumber \\
	&& \qquad \qquad \qquad + e^{-i\varphi} m^2 {\mathcal Y}_{00}^{++}\bigg] |1\rangle \langle 0|.
\end{eqnarray}
The state of the detector is written in terms of the four integrals, 
\begin{equation}
	\label{y00def}
	 \mathcal{Y}_{00}^{\pm \pm}  = \int_{\tau_i}^{\tau_f} \!\!\!\! d\tau' \int_{\tau_i}^{\tau_f} \!\!\!\! d\tau'' \, e^{ i \omega (\pm \tau' \pm \tau'')} \langle \Phi_{0}(\tau') \Phi_{0}(\tau'')\rangle. 
\end{equation}
The subscript `0' on ${\mathcal Y}$ and on the smeared field operators $\Phi$ indicate that the detector is inertial. The matrix element, $\langle \Phi_{0}(\tau') \Phi_{0}(\tau'')\rangle$ is with respect to the Minkowski vacuum $|{\mathbf 0}_M\rangle$ and $ \langle {\mathbf 0}_M| \Phi_{0}(\tau') \Phi_{0}(\tau'')|{\mathbf 0}_M\rangle = {\mathcal W}^{\rm in}_M(\tau',\tau'') $  is given in Eq.~(\ref{wightman3}). Note that for obtaining the state of the detector in Eq.~(\ref{rhofinal0}) we have used $ \langle {\mathbf 0}_M| \Phi_{0}(\tau) |{\mathbf 0}_M\rangle = 0$. In matrix form $\rho_0^S(\tau_f)$ can be written as 
\begin{equation}
	\label{inDM}
	\left[ {\begin{array}{cc}
	\alpha_0 \cos^{2} \frac{\theta}{2} + \beta_0 \sin^{2}\frac{\theta}{2} &  \sin \theta \frac{\kappa_0 e^{-i \varphi} + \lambda_0 e^{i \varphi}}{2} \vphantom{\bigg|} \\ 
	\sin \theta  \frac{\kappa_0^* e^{i \varphi} + \lambda_0^* e^{-i \varphi}}{2} &
	\gamma_0 \sin^{2} \frac{\theta}{2} + \eta_0 \cos^{2} \frac{\theta}{2} \end{array} } \right] 
\end{equation}
with
\begin{eqnarray}
	\label{incoeff}
	\alpha_0 &= &  1 - |m|^2 \Re[{\mathcal T} {\mathcal Y}_{00}^{-+}]  \nonumber \\
	\beta_0 & = & |m|^2  {\mathcal Y}_{00}^{+-} \nonumber \\
	\gamma_0 & = & 1 - |m|^2 \Re[{\mathcal T} {\mathcal Y}_{00}^{+-}] \nonumber \\
	\eta_0 & = & |m|^2  {\mathcal Y}_{00}^{-+} \nonumber \\
	\kappa_0 & = & 1 -\frac{|m|^2}{2} [{\mathcal T} {\mathcal Y}_{00}^{-+}+({\mathcal T} {\mathcal Y}_{00}^{+-})^*] \nonumber \\
	\lambda_0 & = &  (m^*)^2 {\mathcal Y}_{00}^{--}.
\end{eqnarray}
Note that
\begin{eqnarray*}
	{\mathcal T}{\mathcal Y}_{00}^{+-} & =& \int_{\tau_i}^{\tau_f} \!\!\!\! d\tau' \int_{\tau_i}^{\tau'} \!\!\!\! d\tau'' \, e^{ i \omega (\tau' - \tau'')} \langle \Phi_{0}(\tau') \Phi_{0}(\tau'')\rangle \\
	&& + \int_{\tau_i}^{\tau_f} \!\!\!\! d\tau'' \int_{\tau_i}^{\tau''} \!\!\!\! d\tau' \, e^{ i \omega (\tau'' - \tau')} \langle  \Phi_{0}(\tau'') \Phi_{0}(\tau')\rangle.
\end{eqnarray*}
The order of $\tau'$ and $\tau''$ in the exponential in the second term also reverses under the time ordering because the term we are considering arises from a time ordered product of $H_I(\tau')$ and $H_I(\tau'')$. We also have, 
\begin{eqnarray*}
	({\mathcal T}{\mathcal Y}_{00}^{+-})^* & =& \int_{\tau_i}^{\tau_f} \!\!\!\! d\tau' \int_{\tau_i}^{\tau'} \!\!\!\! d\tau'' \, e^{ i \omega (\tau'' - \tau')} \langle \Phi_{0}(\tau'') \Phi_{0}(\tau')\rangle \\
	&& + \int_{\tau_i}^{\tau_f} \!\!\!\! d\tau'' \int_{\tau_i}^{\tau''} \!\!\!\! d\tau' \, e^{ i \omega (\tau' - \tau'')} \langle  \Phi_{0}(\tau') \Phi_{0}(\tau'')\rangle
\end{eqnarray*}
The first term of ${\mathcal T}{\mathcal Y}_{00}^{+-}$ is a restricted integral over half of a square of length $\tau_f - \tau_i$ in the $(\tau',\tau'')$ plane while the second term of $({\mathcal T}{\mathcal Y}_{00}^{+-})^*$ is the integral over the the remaining half with the same integrand. The same applies to the second term of ${\mathcal T}{\mathcal Y}_{00}^{+-}$ and first term of $({\mathcal T}{\mathcal Y}_{00}^{+-})^*$ with the integrands of the two pairs of terms being complex conjugates of each other. Using this observation, we obtain,
\begin{eqnarray*}
	\Re[{\mathcal T}{\mathcal Y}_{00}^{+-}]& \!\! = \! &\frac{1}{2} \bigg[ \int_{\tau_i}^{\tau_f} \!\!\!\! d\tau' \int_{\tau_i}^{\tau_f} \!\!\!\! d\tau'' \, e^{ i \omega (\tau' - \tau'')} \langle  \Phi_{0}(\tau') \Phi_{0}(\tau'')\rangle \nonumber \\  
	&& + \int_{\tau_i}^{\tau_f} \!\!\!\! d\tau' \int_{\tau_i}^{\tau_f} \!\!\!\! d\tau'' e^{ i \omega (\tau'' - \tau')} \langle  \Phi_{0}(\tau'') \Phi_{0}(\tau') \rangle \bigg]  \nonumber \\ 
	&\!\! = \!& \int_{\tau_i}^{\tau_f} \!\!\!\!\! d\tau'  \!\!\!\int_{\tau_i}^{\tau_f} \!\!\!\!\! d\tau'' \Re\Big[ e^{ i \omega (\tau'' - \tau')} \langle  \Phi_{0}(\tau'') \Phi_{0}(\tau') \rangle \Big]  \nonumber \\
	& \!\! = \! & \Re[{\mathcal Y}_{00}^{+-}] .
\end{eqnarray*} 
Let us now write ${\mathcal Y}_{00}^{+-}$ as 
\begin{eqnarray*}
	{\mathcal Y}_{00}^{+-} & =& \int_{\tau_i}^{\tau_f} \!\!\!\! d\tau' \int_{\tau_i}^{\tau'} \!\!\!\! d\tau'' \, e^{ i \omega (\tau' - \tau'')} \langle \Phi_{0}(\tau') \Phi_{0}(\tau'')\rangle \\
	&& + \int_{\tau_i}^{\tau_f} \!\!\!\! d\tau'' \int_{\tau_i}^{\tau''} \!\!\!\! d\tau' \, e^{ i \omega (\tau' - \tau'')} \langle  \Phi_{0}(\tau') \Phi_{0}(\tau'')\rangle,
\end{eqnarray*}
where we have split the integration regions into two halves, one in which $\tau' \geq \tau''$ and the other in which $\tau'' > \tau'$. By interchanging $\tau'$ and $\tau''$ in the second term in the equation above we obtain, 
\begin{eqnarray*}
	{\mathcal Y}_{00}^{+-} & =&  \int_{\tau_i}^{\tau_f} \!\!\!\! d\tau' \int_{\tau_i}^{\tau'} \!\!\!\! d\tau'' \, \Re \Big[e^{ i \omega (\tau' - \tau'')} \langle \Phi_{0}(\tau') \Phi_{0}(\tau'')\rangle \Big].
\end{eqnarray*}
This means that ${\mathcal Y}_{00}^{+-}$ is real and therefore $\Re[{\mathcal T}{\mathcal Y}_{00}^{+-}] = {\mathcal Y}_{00}^{+-}$.  Similarly we can show that $\Re[{\mathcal T}{\mathcal Y}_{00}^{-+}]={\mathcal Y}_{00}^{-+}$. Using these results we find that $\alpha_0 + \eta_0 = \beta_0 + \gamma_0=1$ and $ {\rm tr}[\rho_0^S(\tau_f)] = 1$. The trace of the reduced density matrix of the detector is preserved by the open dynamics even if the joint system-environment unitary time evolution operator is truncated at second order in Eq.~(\ref{unitary1}). 

\subsection{Uniformly accelerated detector}

Calculation of the reduced state of a detector that is uniformly accelerated with acceleration $a$ in the interval $[\tau_i, \tau_f]$ proceeds exactly as in the case of the inertial detector. The only change is that the field operators are evaluated along the accelerated trajectory in 1+1 dimensions given by 
\begin{equation}
	\label{accelcoord}
	x_\tau =  \frac{1}{a}\cosh(a \tau) -  \frac{1}{a} \quad {\rm and} \quad 
	t_\tau =   \frac{1}{a} \sinh(a \tau).
\end{equation}
The state of the uniformly accelerated detector at $\tau_f$ is given by the same expressions in Eqs.~(\ref{inDM}) and (\ref{incoeff}) with the simple replacement of the subscript `0' with `$a$' to indicate accelerated motion. The relevant integrals appearing in the new expressions for the accelerated detector are,
\begin{equation}
	\label{yaadef}
	 \mathcal{Y}_{aa}^{\pm \pm}  = \int_{\tau_i}^{\tau_f} \!\!\!\! d\tau' \int_{\tau_i}^{\tau_f} \!\!\!\! d\tau'' \, e^{ i \omega (\pm \tau' \pm \tau'')} \langle \Phi_{a}(\tau') \Phi_{a}(\tau'')\rangle. 
\end{equation}
The Wightman function for the finite size detector appearing in these integrals is
\begin{eqnarray*}
	{\mathcal W}_M^a(\tau',\tau'') & =& -\frac{1}{4 \pi^2} \Big\{ [t_{\tau'} - t_{\tau''} -i\epsilon(\dot{t}_{\tau'} + \dot{t}_{\tau''})]^2 \\
	&& \qquad  - [x_{\tau'} - x_{\tau''} -i \epsilon(\dot{x}_{\tau'} + \dot{x}_{\tau''})]^2 \Big\}^{-1}
\end{eqnarray*}
Using Eq.~(\ref{accelcoord}) in the expression above, we obtain,
\begin{equation*}
	{\mathcal W}_M^a(\tau',\tau'') = \frac{-1/16 \pi^{2}}{\Big[ \frac{1}{a}\sinh(a \frac{\tau' - \tau''}{2}) - i\epsilon\cosh(a \frac{\tau' - \tau''}{2})\Big]^{2}}.
\end{equation*}
The term $\eta_{a} = |m|^2 {\mathcal Y}_{aa}^{-+} $ is the well known  amplitude for  the accelerated detector that is initially in its ground state transitioning to its excited state after interacting with the vacuum field~\cite{ApplicationsUnruh}.

\section{Inertial followed by accelerated motion \label{InAc}}

We now  let the detector be at rest (inertial) in the lab frame from $\tau_{0i} = -{\mathtt t}$ to   $\tau_{0f} = 0$ followed by constant uniform acceleration at rate $a$ from $\tau_{ai} = 0$ to $\tau_{af}={\mathtt T}$. We are  interested in the open dynamics of the detector during its trajectory from $t=0$ to ${\mathtt T}$. We already know how to compute the state of the detector at $t=0$ following a period of inertial motion. Here we compute the state at $t={\mathtt T}$ directly by evolving the initial state $|\psi_i\rangle |{\mathbf 0}_M\rangle$ at $t=-{\mathtt t}$ with the unitary operator
\begin{eqnarray}
	\label{unitary2}
	U({\mathtt T}, -{\mathtt t}) & = & \openone - i \int_{-{\mathtt t}}^0 d\tau' H_{0I}(\tau') -i \int_0^{\mathtt T} d\tau' H_{aI}(\tau') \nonumber \\
	&& \quad - \frac{1}{2} {\mathcal T} \int_{-{\mathtt t}}^0 \!\!\!\! d\tau'\int_{-{\mathtt t}}^0 \!\!\!\! d\tau''\, H_{0I}(\tau')H_{0I}(\tau'') \nonumber \\
	&& \quad - \frac{1}{2} {\mathcal T} \int_0^{\mathtt T} \!\!\!\! d\tau'\int_0^{\mathtt T} \!\!\!\! d\tau'' H_{aI}(\tau')H_{aI}(\tau'') \nonumber \\
	&& \quad - \int_0^{\mathtt T} \!\!\!\! d\tau'\int_{-{\mathtt t}}^{0} \!\!\!\! d\tau'' H_{aI}(\tau')H_{0I}(\tau''),
\end{eqnarray}
where, following Eq.~(\ref{Hinteraction}) we define $H_{0I}(\tau)=\hat{m}(\tau) \Phi_0(\tau)$ and $H_{aI}(\tau)=\hat{m}(\tau) \Phi_a(\tau)$. Using this we compute,
\begin{eqnarray}
    \label{finalstate2}
	|\Psi_{0a}^f\rangle & \! \!=  \! \! &  U({\mathtt T}, -{\mathtt t}) |\psi_i\rangle |{\mathbf 0}_M\rangle  \nonumber \\
	& \! \! = \! \! & |0\rangle \bigg\{ \cos \frac{\theta}{2} \bigg[ \openone -\frac{|m|^2}{2} \big[{\mathcal T} \hat{A}_0^-(-{\mathtt t}, 0) + {\mathcal T} \hat{A}_a^-(0, {\mathtt T})] \nonumber \\
	&& \qquad  + \, |m|^2 \hat{A}_{0a}^-(-{\mathtt t}, {\mathtt T})\bigg] |{\mathbf 0}_M\rangle  \nonumber \\
	&& \qquad  -im^* \sin \frac{\theta}{2}e^{i\varphi} \big[ \hat{B}_0^\dagger(-{\mathtt t},0)+\hat{B}_a^\dagger(0,{\mathtt T})\big]|{\mathbf 0}_M\rangle \bigg\} \nonumber \\
	&& +|1\rangle \bigg\{ \!\!\sin \frac{\theta}{2} e^{i\varphi} \bigg[\! \openone \!- \!\frac{|m|^2}{2} \big[{\mathcal T}\! \hat{A}_0^+(-{\mathtt t}, 0) \! + \! {\mathcal T}\! \hat{A}_a^+(0, {\mathtt T})] \nonumber \\
	&& \qquad  + \, |m|^2 \hat{A}_{0a}^+(-{\mathtt t}, {\mathtt T})\bigg] |{\mathbf 0}_M\rangle  \nonumber \\
	&& \qquad  -im \cos \frac{\theta}{2} \big[ \hat{B}_0(-{\mathtt t},0)\!+\!\hat{B}_a(0,{\mathtt T})\big]|{\mathbf 0}_M\rangle \!\bigg\}.
\end{eqnarray}
We have extended the definitions of the operators $\hat{A}$ and $\hat{B}$ in Eq.~(\ref{ABdef}) in an obvious manner to the trajectory considered here and introduced subscripts denoting inertial, accelerated and inertial followed by accelerated motion. Using Eq.~(\ref{finalstate2}) we have to compute, 
\begin{equation*}
      \rho^S_{0a}(T) =  \Tr_{E}\!\!\big[ U({\mathtt T}, -{\mathtt t}) |\psi_i\rangle |{\mathbf 0}_M\rangle \langle {\mathbf 0}_M|\langle \psi_i| U({\mathtt T}, -{\mathtt t}) \big],
\end{equation*}
We see from a comparison of the equation above with Eqs.~(\ref{psif0}) and (\ref{psif1}) that $\rho^S_{0a}(T)$ can be written in exactly the same form as $\rho_0^S(\tau_f)$ given in Eq.~(\ref{inDM}) with modified coefficients given by
\begin{eqnarray}
	\label{mixedcoeff}
	\alpha_{0a} &\!\!= \!\!& 1 - |m|^2\big( \Re[{\mathcal T}{\mathcal Y}_{00}^{-+}] + \Re[{\mathcal T}{\mathcal Y}_{aa}^{-+}]+{\mathcal Y}_{0a}^{-+}\! + \!{\mathcal Y}_{a0}^{-+}] \big) \nonumber \\
	\beta_{0a} & \!\!= \!\! & |m|^2\big( {\mathcal Y}_{00}^{+-} + {\mathcal Y}_{aa}^{+-} + {\mathcal Y}_{0a}^{+-} + {\mathcal Y}_{a0}^{+-} \big) \nonumber \\
	\gamma_{0a} & \!\!= \!\!& 1 - |m|^2\big( \Re[{\mathcal T}{\mathcal Y}_{00}^{+-}] + \Re[{\mathcal T}{\mathcal Y}_{aa}^{+-}]+{\mathcal Y}_{0a}^{+-}\! + \!{\mathcal Y}_{a0}^{+-}] \big) \nonumber \\
	\eta_{0a} & \!\!= \!\! & |m|^2\big( {\mathcal Y}_{00}^{-+} + {\mathcal Y}_{aa}^{-+} + {\mathcal Y}_{0a}^{-+} + {\mathcal Y}_{a0}^{-+} \big) \nonumber \\
	\kappa_{0a} & \!\!= \!\! & 1- \frac{|m|^2}{2} \big[ {\mathcal T}{\mathcal Y}_{00}^{-+} + ({\mathcal T}{\mathcal Y}_{00}^{+-})^*  +  {\mathcal T}{\mathcal Y}_{aa}^{-+} + ({\mathcal T}{\mathcal Y}_{aa}^{+-})^*\nonumber \\
	&& \qquad \qquad + \, 2{\mathcal Y}_{a0}^{-+}+ 2{\mathcal Y}_{0a}^{+-} \big] \nonumber \\
	\lambda_{0a} & \!\!= \!\! & (m^*)^2 \big[ {\mathcal Y}_{00}^{--} + {\mathcal Y}_{aa}^{--} + {\mathcal Y}_{0a}^{--} + {\mathcal Y}_{a0}^{--} \big].
\end{eqnarray}
Here we have introduced eight more new integrals,
\[ \mathcal{Y}_{0a}^{\pm \pm}  = \int_{\tau_i}^{\tau_f} \!\!\!\! d\tau' \int_{\tau_i}^{\tau_f} \!\!\!\! d\tau'' \, e^{ i \omega (\pm \tau' \pm \tau'')} \langle \Phi_{0}(\tau') \Phi_{a}(\tau'')\rangle. \]
and
\[ \mathcal{Y}_{a0}^{\pm \pm}  = \int_{\tau_i}^{\tau_f} \!\!\!\! d\tau' \int_{\tau_i}^{\tau_f} \!\!\!\! d\tau'' \, e^{ i \omega (\pm \tau' \pm \tau'')} \langle \Phi_{a}(\tau') \Phi_{0}(\tau'')\rangle. \]
Note that $\mathcal{Y}_{0a}^{\pm \mp} = (\mathcal{Y}_{a0}^{\mp \pm})^{*}$,  $\mathcal{Y}_{0a}^{+ +} = (\mathcal{Y}_{a0}^{--})^{*}$ and  $\mathcal{Y}_{0a}^{- -} = (\mathcal{Y}_{a0}^{++})^{*}$. The relevant Wightman functions ${\mathcal W}_M^{0a}(\tau', \tau'')$ and ${\mathcal W}_M^{a0}(\tau', \tau'')$ appearing in these integrals are 
\begin{eqnarray}
\label{cross_Wightman}
	\langle \Phi_{0}(\tau') \Phi_{a}(\tau'')\rangle & \! \! = \! \! & -\frac{1}{4 \pi^2} \bigg\{ \bigg[\tau' - \frac{1}{a} \sinh(a \tau'') \nonumber \\
	&& \qquad \qquad \qquad    - i\epsilon\big( 1 + \cosh(a\tau'') \big)\bigg]^{2} \nonumber \\
	&&  -\bigg[ \frac{1}{a}\cosh(a\tau'') \! -\! \frac{1}{a} \! + \! i \epsilon \sinh(a\tau'')  \bigg]^{2} \bigg\}^{-1}\!\!\!\!, \nonumber \\
	\langle \Phi_{0}(\tau') \Phi_{a}(\tau'')\rangle & \! \! = \! \!  & -\frac{1}{4 \pi^2} \bigg\{ \bigg[\frac{1}{a} \sinh(a \tau') - \tau'' \nonumber \\
	&& \qquad \qquad \qquad    - i\epsilon\big(\cosh(a\tau') + 1 \big)\bigg]^{2} \nonumber \\
	&&  -\bigg[ \frac{1}{a}\cosh(a\tau')\! -\! \frac{1}{a}\! -\! i \epsilon \sinh(a\tau')  \bigg]^{2} \bigg\}^{-1}\!\!\!\! . \nonumber \\
\end{eqnarray}

Using $\Re[{\mathcal T}{\mathcal Y}_{00}^{-+}] = {\mathcal Y}_{00}^{-+}$, $\Re[{\mathcal T}{\mathcal Y}_{00}^{+-}] = {\mathcal Y}_{00}^{+-}$, $\Re[{\mathcal T}{\mathcal Y}_{aa}^{-+}] = {\mathcal Y}_{aa}^{-+}$ and, $\Re[{\mathcal T}{\mathcal Y}_{aa}^{+-}] = {\mathcal Y}_{aa}^{+-}$ in Eq.~(\ref{mixedcoeff}) we find that $\alpha_{0a}+\eta_{0a}=\gamma_{0a} + \beta_{0a} = 1$ and hence ${\rm tr}[\rho^S_{0a}(T)]=1$ for all values of $-{\mathtt t}$ and ${\mathtt T}$. 

\section{Dynamical maps \label{dynmap}}

Dynamical maps furnish a mathematical description of the finite-time dynamics of an open quantum system. They capture the effect of the environment on the open system in addition to its unitary, closed evolution. Dynamical maps are super-operators that admits a matrix representation of the form, 
\begin{equation}\label{A transform}
    \rho_{r,s}(\tau_f) = A_{rs,r's'}(\tau_i,\tau_{f}) \rho_{r's'}(\tau_{i})
\end{equation}
In order to construct this matrix representation we assume that the density matrix of the system has been `vectorised' \cite{bengtsson_zyczkowski_2006} by casting it as a column vector. For instance, the density matrices we have considered for the detector like the one in Eq.~(\ref{inDM}) can be written in vectorised form as 
\[ \vec{\rho}^S_j(\tau_f) = \left( \begin{array}{c}
	\alpha_j \cos^{2} \frac{\theta}{2} + \beta_j \sin^{2}\frac{\theta}{2}  \\
	  \sin \theta \frac{\kappa_j e^{-i \varphi} + \lambda_j e^{i \varphi}}{2}  \\ 
	\sin \theta  \frac{\kappa_j^* e^{i \varphi} + \lambda_j^* e^{-i \varphi}}{2} \\
	\gamma_j \sin^{2} \frac{\theta}{2} + \eta_j \cos^{2} \frac{\theta}{2},
\end{array} \right) \]
where $j=0,a,0a$. At this point it is convenient to add another value for the subscript $j$ as $j={\rm ini}$ with $\vec{\rho}_{\rm ini}^S(\tau)$ indicating the state of the detector, $|\psi_i\rangle \langle \psi_i|$ at the time of switch on. For this state we have $\alpha_{\rm ini} = \gamma_{\rm ini} = \kappa_{\rm ini} = 1$ and $\beta_{\rm ini} = \lambda_{\rm ini} = \eta_{\rm ini} = 0$. With this definition we see that all the map matrices we have to find are those satisfying an equation of the form,
\begin{equation}
	\label{Amap1}
	\vec{\rho}^S_j(\tau_f) = A^{(j,k)}\vec{\rho}_k^S(\tau_i).
\end{equation}
By inspection we see that the $4 \times 4$ matrix $A^{(j,k)}$ has eight non-zero elements which satisfy the following equations,
\begin{eqnarray*}
	A^{(j,k)}_{11} \alpha_k + A^{(j,k)}_{14} \eta_k  = \alpha_j, &\quad & 
	A^{(j,k)}_{11} \beta_k + A^{(j,k)}_{14} \gamma_k =  \beta_j, \nonumber \\
	A^{(j,k)}_{22} \kappa_k + A^{(j,k)}_{23} \lambda_k^*  =  \kappa_j, & \quad & 
	A^{(j,k)}_{22} \lambda_k + A^{(j,k)}_{23} \kappa_k^*  =  \lambda_j, \nonumber \\
	A^{(j,k)}_{32} \kappa_k + A^{(j,k)}_{33} \lambda_k^*  =  \lambda_j^*, & \quad & 
	A^{(j,k)}_{32} \lambda_k + A^{(j,k)}_{33} \kappa_k^*  =  \kappa_j^*, \nonumber \\
	A^{(j,k)}_{41} \alpha_k + A^{(j,k)}_{44} \eta_k =  \eta_j, & \quad & 
	A^{(j,k)}_{41} \beta_k + A^{(j,k)}_{44} \gamma_k  =  \gamma_j, 
\end{eqnarray*}
with solutions,
\begin{eqnarray}
	\label{Amapelements}
	A^{(j,k)}_{11} = \frac{\alpha_j \gamma_k - \beta_j \eta_k}{\alpha_k \gamma_k - \beta_k \eta_k}, & \quad & 
	A^{(j,k)}_{14}  = \frac{\beta_j \alpha_k - \alpha_j \beta_k}{\alpha_k \gamma_k - \beta_k \eta_k}, \nonumber \\
	A^{(j,k)}_{22} \! = \! \frac{\kappa_j \kappa_k^{*} - \lambda_j (\lambda_k)^{*}}{|\kappa_k|^{2} - |\lambda_k|^{2}}, & \quad & 
	A^{(j,k)}_{23}  = \frac{\lambda_j \kappa_k - \kappa_j \lambda_k}{|\kappa_k|^{2} - |\lambda_k|^{2}} , \nonumber \\
	A^{(j,k)}_{32} = \frac{\lambda_j^{*} \kappa_k^{*} - \kappa_j^{*} \lambda_k^{*}}{|\kappa_k|^{2} - |\lambda_k|^{2}}, & \quad & 
	A^{(j,k)}_{33} = \frac{\kappa_j^{*} \kappa_k - \lambda_j^{*} \lambda_k}{|\kappa_k|^{2} - |\lambda_k|^{2}}, \nonumber  \\
	A^{(j,k)}_{41} = \frac{\eta_j \gamma_k - \gamma_j \eta_k}{\alpha_k \gamma_k - \beta_k \eta_k}, & \quad & 
	A^{(j,k)}_{44} = \frac{\gamma_j \alpha_k - \eta_j \beta_k}{\alpha_k \gamma_k - \beta_k \eta_k }. \quad \;\;\;\;
\end{eqnarray}

It is significant to note here that none of the matrix elements above depend on the parameters $\theta$ and $\varphi$ of the initial state of the detector. This is essential to ensure that the dynamical map is linear. These matrix elements depend on the duration of inertial and accelerated motions, the energy difference $\omega$ between the two levels of the detector and its size $\epsilon$. 

\subsection{Complete Positivity and Positivity}\label{NCP Maps}

The matrix representation of the dynamical map we have obtained in terms of the $A$-matrix is not Hermitian. A Hermitian representation can be obtained by re-shuffling the $A$-matrix to the so-called $B$-matrix form~\cite{ECGPhysRev.121.920} as
\begin{equation}
    B_{rr',ss'} = A_{rs,r's'}.
\end{equation}
The eigenvalues and eigenvectors of the $B$-matrix lead to well known operator-sum form for completely positive dynamical maps, 
\[ \rho^S(\tau_f) = \sum_j K_j \rho^S(\tau_i)K_j^\dagger, \qquad \sum_j K_j^\dagger K_j = \openone.\]
Since we are interested only in the nature of the dynamical map induced by the motion of the detector, it suffices to find the eigenvalues of the $B$-matrix. If they are all positive then it corresponds to a completely positive map with an operator-sum form as given above~\cite{bengtsson_zyczkowski_2006,choi_1972,Quanta77,nielsen_chuang_2010}. If one or more of its eigenvalues are not positive then the dynamical map may be positivity preserving in some cases and in others its action may even be negative on part of the state space of the system. Lack of complete positivity, or for that matter, lack of positivity of dynamical maps is a consequence of initial correlations between the system and environment~\cite{ShajiInitialEntangled} and for the detector such correlations arise because of the unavoidable vacuum fluctuations of the field. 

\subsubsection{Inertial or accelerated motion}

\begin{figure}[!htb]
	\resizebox{8.5cm}{5.5cm}{\includegraphics{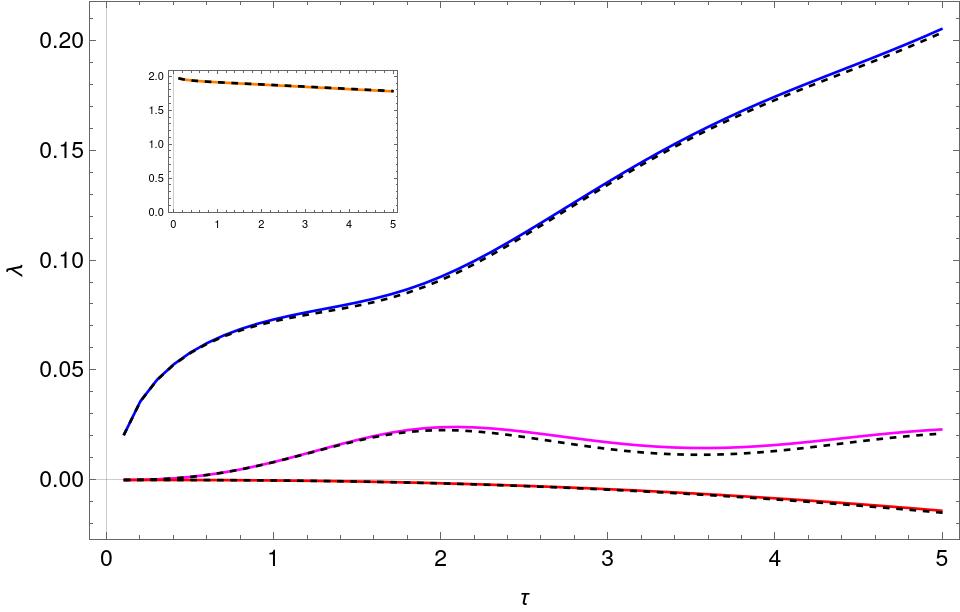}}
	\caption{Solid lines are the eigenvalues of $B_a$ as a function of the duration of uniformly accelerated motion while the dashed lines that follow the solid-lines closely correspond to the eigenvalues of $B_0$ for a detector in inertial motion for the same time period. The largest eigenvalue in both cases is plotted separately in the inset for clarity since it is much larger in magnitude compared to the others.}
	\label{fig:1}
\end{figure}

The $A$-matrix corresponding to inertial motion during the interval $(-{\mathtt t},0)$ is obtained from Eq.~(\ref{Amap1}) by setting $k={\rm ini}$ and $j=0$ with $\tau_i= -{\mathtt t}$ and $\tau_f = 0$. After re-arrangement, we obtain, 
\begin{equation}
	\label{Binertial}
	B_0 = \left( \begin{array}{cccc}
		\alpha_0 & 0 & 0 & \kappa_0 \\
		0 & \beta_0 & \lambda_0 & 0 \\
		0 & \lambda_0^* & \eta_0 & 0 \\
		\kappa_0^* & 0 & 0 & \gamma_0
	\end{array} \right).
\end{equation}
The eigenvalues of $B_0$-matrix are plotted as dashed lines in Fig.~\ref{fig:1} as a function of the duration of inertial motion $\tau = |-{\mathtt t}|$. We see that three of the four eigenvalues of $B_0$ are always positive. However since the initial state $\rho^S_{\rm ini}(-{\mathtt t})$ is assumed to be uncorrelated with the field at the time of switch-on of the detector we expect $B_0$ to be completely positive with all its eigenvalues being positive as well. The slight negative value taken by the smallest eigenvalue of $B_0$ is because of the truncation to the second order term of the exponential in the unitary time evolution operator. Keeping this limitation in mind, we consider an induced dynamical map to be not completely positive only if more than one of its eigenvalues become negative. 

The computation of $B_a$ which corresponds to a detector that is uniformly accelerated also proceeds in an identical manner with the $B$-matrix having exactly the same form as in Eq.~(\ref{Binertial}) with subscript $0$ replaced by $a$ in the matrix elements. The eigenvalues of $B_a$ are also plotted in Fig.~\ref{fig:1} as solid lines. We see that the eigenvalues of the $B$-matrix for the inertial and accelerated cases are quite close to each other, which highlights the practical difficulties of observing these effects unless the acceleration is very large.

\subsubsection{Inertial followed by accelerated motion}

\begin{figure}[!htb]
    \resizebox{8.5cm}{5.5cm}{\includegraphics{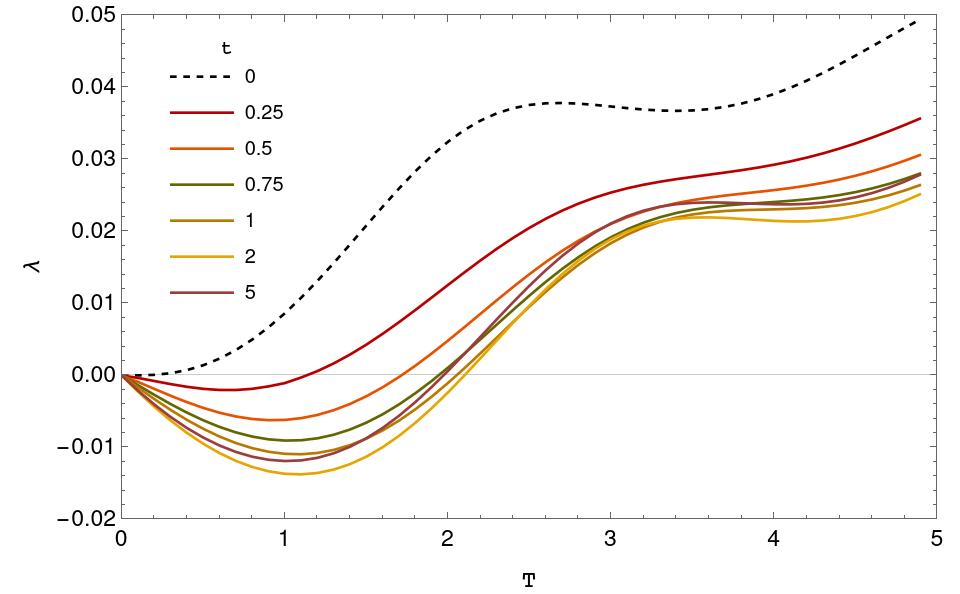}}
    \caption{The second-smallest eigenvalue of the dynamical matrix $B_{0a}$ is plotted as a function of time for different durations, ${\mathtt t}$, of inertial motion prior to accelerated motion with  $a=3$. ${\mathtt T}$ represents the duration for which the detector is accelerated. The dashed line represents the case where there is no initial period of time during which the detector is inertial. We see that the initial period of inertial motion leads to dynamics that is not completely positive, indicating entanglement that is built up between the detector and field during this period. For all the plots, we have used $\omega = 1$ and $\epsilon = 0.025$.}
    \label{fig:2a}
\end{figure}

\begin{figure}[!htb]
    \resizebox{8.5cm}{5.5cm}{\includegraphics{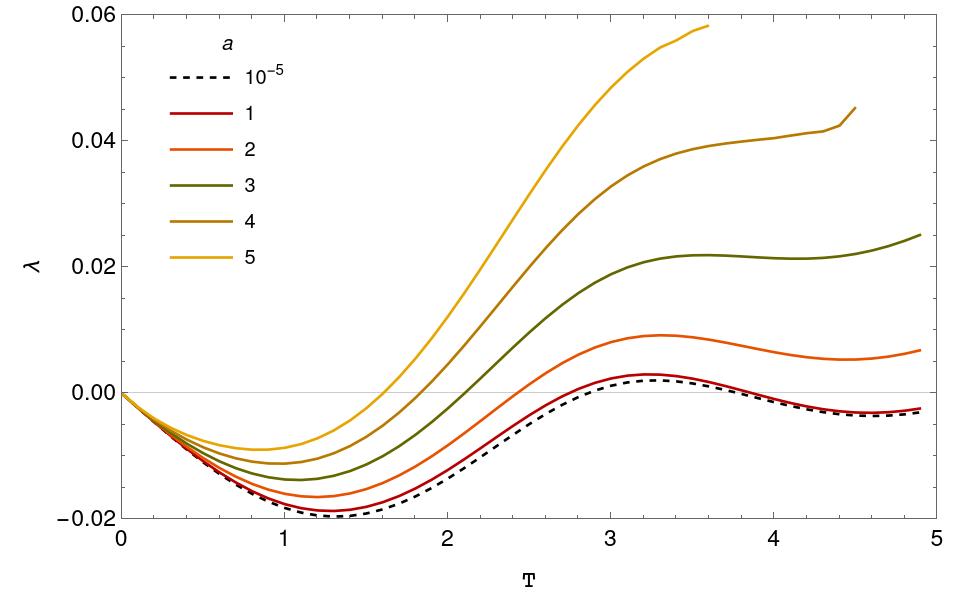}}
    \caption{The second-smallest eigenvalue of the dynamical matrix $B_{0a}$ is plotted as a function of time for different values of acceleration $a$, keeping the duration of initial inertial motion constant at ${\mathtt t}=1$. The dashed line represents the case with $a=0$. We see that the initial period of inertial motion leads to NCP dynamics irrespective of the acceleration. The duration for which the dynamics is NCP reduces with increasing acceleration. For all the plots, $\omega = 1$ and $\epsilon = 0.025$.}
    \label{fig:2b}
\end{figure}

We now consider the detector initialised in the state $\rho^S_{\rm ini}(-{\mathtt t})$ evolving to $\rho^S_0(0)$ after inertial motion for a duration ${\mathtt t}$  and subsequently accelerated uniformly so that its final state is $\rho^S_{0a}({\mathtt T})$. We are interested in the dynamical map connecting $\rho^S_0(0)$ and $\rho^S_{0a}({\mathtt T})$ whose elements can be obtained using Eqs.~(\ref{Amapelements}), (\ref{incoeff}) and, (\ref{mixedcoeff}). The second-smallest eigenvalue of the rearranged map matrix, $B_{0a}$, is plotted in Figs.~\ref{fig:2a} and \ref{fig:2b} as a function of time for various cases. We choose to show the second-smallest rather than the smallest to avoid the negative eigenvalue produced due to the truncation of $U(\tau_{f},\tau_{i})$. 

In Fig.~\ref{fig:2a}, dependence of the nature of the dynamical map on the duration of initial, inertial motion is shown. We first verify that when this duration is zero, the second-smallest eigenvalue of $B_{0a}$ does remain positive irrespective of the duration of accelerated motion, validating our choice of this eigenvalue as the test of not completely positive dynamics. The smallest eigenvalue in all these cases follows the same behaviour as in the case with only inertial or accelerated motion. From Fig.~\ref{fig:2a} we see that as the period of inertial motion is increased, the duration for which the dynamics remains NCP increases. However this trend does not continue and when ${\mathtt t}$ is greater than 2, the trend reverses. This is consistent with the fact that when ${\mathtt t} \rightarrow \infty$, the probability that the detector is excited due to its interaction with the vacuum field goes to zero.  

In Fig.~\ref{fig:2b}, the acceleration, $a$, is varied keeping the duration of inertial motion constant. For larger values of $a$ the dynamics is NCP for smaller durations. The observed trend is that when $a\rightarrow \infty$ the dynamics again becomes completely positive. We cannot compute the behaviour of the dynamics for very large values of $a$ due to numerical instabilities that appear in the integration routines required to compute the matrix elements of the dynamical map, however the computable values show a clear trend. The dotted line in Fig.~\ref{fig:2b} shows the behaviour of the second-smallest eigenvalue when the acceleration is zero. Here again, the dynamical map is NCP because we are dividing the inertial evolution between $-{\mathtt t}$ and ${\mathtt T}$ into two parts and looking at the dynamical map corresponding to the second part of the evolution. In this case, there exists a single, well-defined dynamical map that connects $\rho^S_{\rm ini}(-{\mathtt t})$ with $\rho^S_{00}({\mathtt T})$ ($0a \rightarrow 00$ for $a=0$) that is completely positive and there is no physical reason for splitting this evolution into two parts since the motion is inertial for the entire duration.

The second-smallest eigenvalue being negative only shows that the dynamical map induced by accelerated motion after is NCP. However, it does not reveal whether the map is positive or not. Typically, with initial entanglement or other quantum correlations between the system and its environment, the reduced dynamics of the system is not even positive~\cite{ShajiInitialEntangled}. The resolution of the question of how physical states can be mapped to nonphysical (negative) ones by the dynamics lies in the observation that with initial entanglement between the system and the environment, certain initial states of the system are not physically possible. Application of the dynamical map blindly to these states can lead to nonphysical results while the action of the map on all states of the system that are consistent with the specification of the initial system-environment correlations will be positivity preserving~\cite{ShajiInitialEntangled,joseph_reference_2018}. 

\begin{figure}[!htb]
	\resizebox{4.2cm}{4.2cm}{\includegraphics{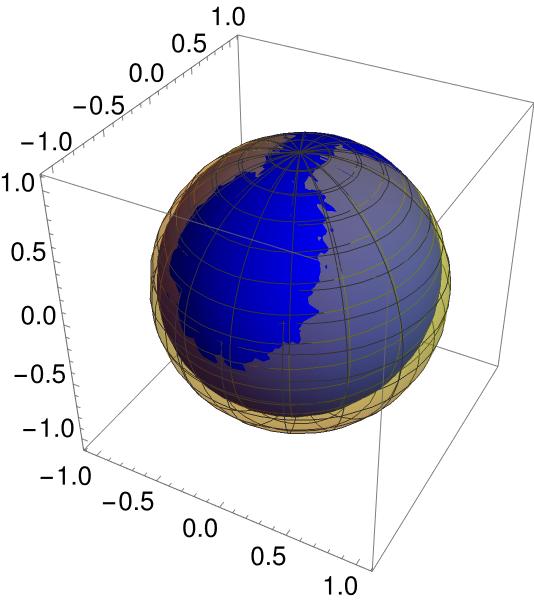}}
	\resizebox{4.2cm}{4.2cm}{\includegraphics{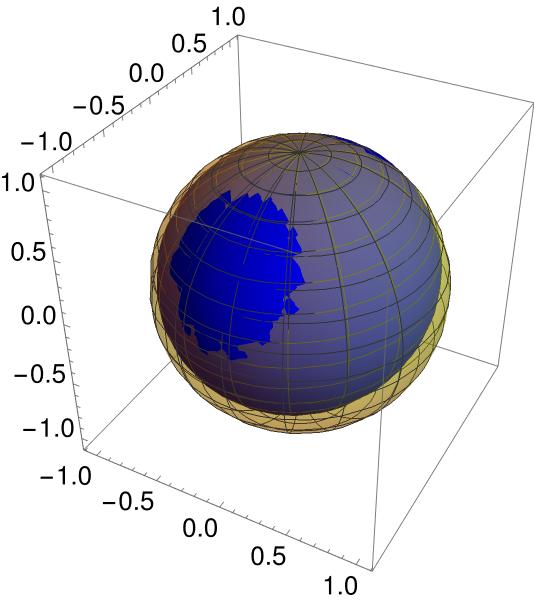}}
	\resizebox{4.2cm}{4.2cm}{\includegraphics{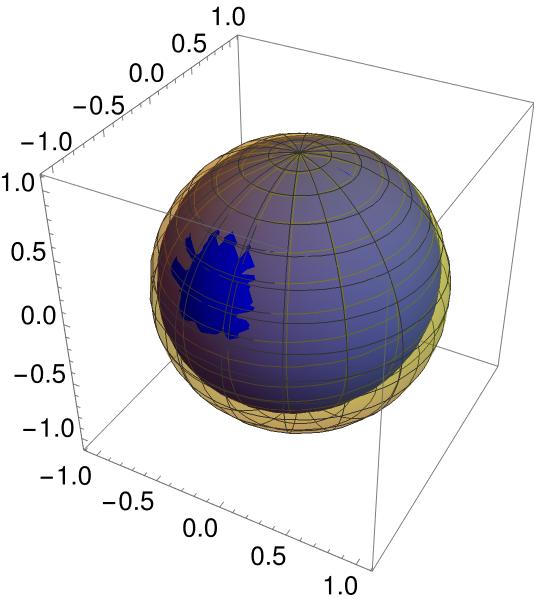}}
	\resizebox{4.2cm}{4.2cm}{\includegraphics{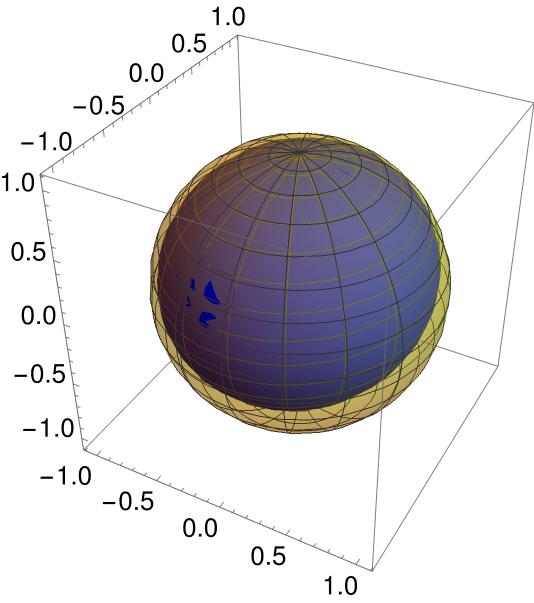}}
	\caption{ TThe Bloch sphere which represents the full state space of the detector is shown in translucent yellow colour in the four figures above. The set of states into which the dynamical map $B_{0a}$ maps the Bloch sphere of states is shown in blue. The parts of this set that protrudes outside the Bloch sphere correspond to unphysical states. The four figures, seen clockwise from top-left, correspond to $a=1$, $2$, $3$ and $4$ respectively with $-{\mathtt t} = -2$, ${\mathtt T} =2$, $\omega = 1$ and $\epsilon = 0.025$.  \label{fig4}}	
\end{figure}

\begin{figure}[!htb]
	\resizebox{4.2cm}{4.2cm}{\includegraphics{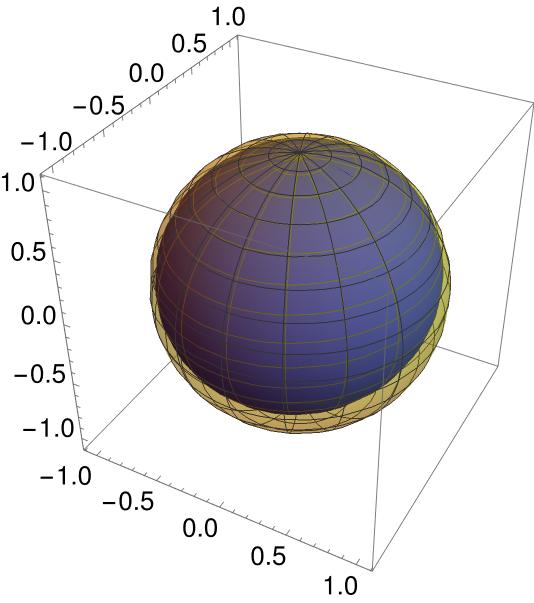}}
	\resizebox{4.2cm}{4.2cm}{\includegraphics{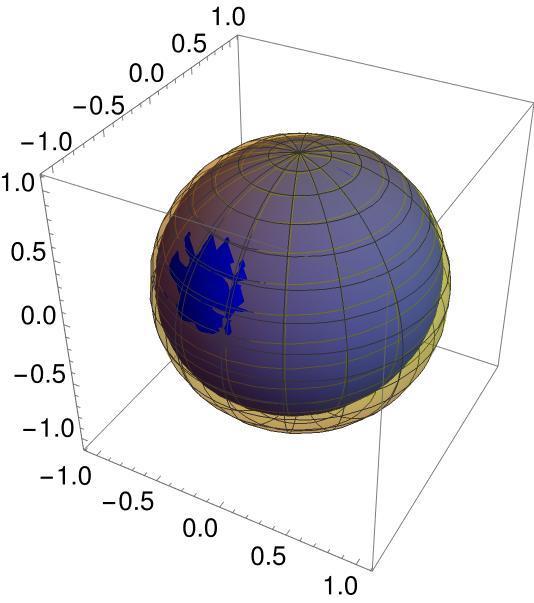}}
	\resizebox{4.2cm}{4.2cm}{\includegraphics{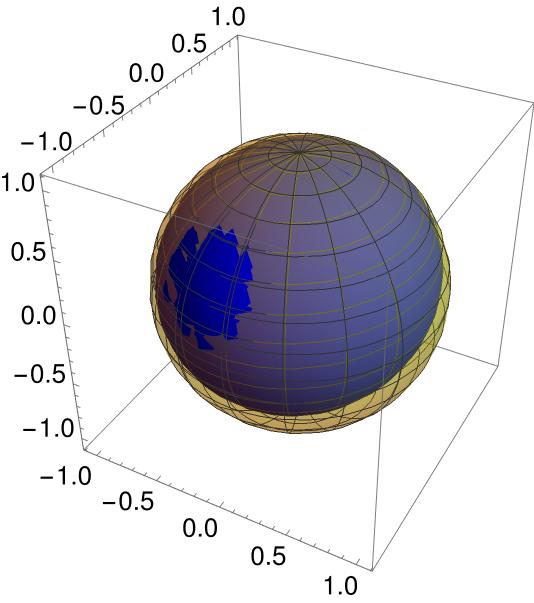}}
	\resizebox{4.2cm}{4.2cm}{\includegraphics{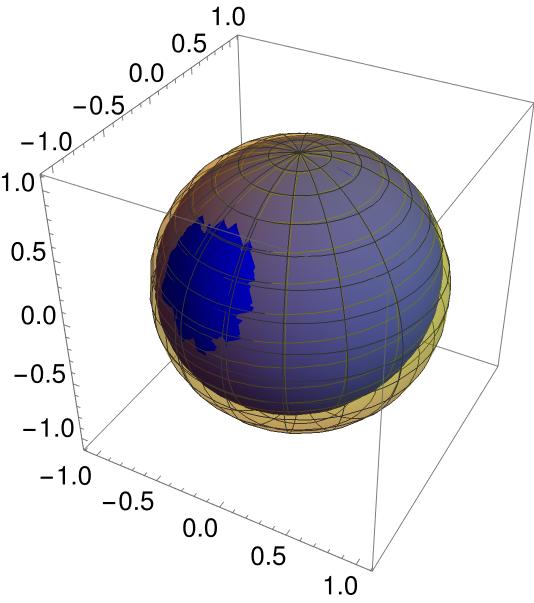}}
	\caption{ The Bloch sphere which represents the full state space of the detector is shown in translucent yellow colour in the four figures above. The set of states into which the dynamical map $B_{0a}$ maps the Bloch sphere of states is shown in blue. The parts of this set that protrudes outside the Bloch sphere correspond to unphysical states. The four figures, seen clockwise from top-left, correspond to ${\mathtt t}=0.25$, $0.75$, $1$ and $2$ respectively with $a = 3$, ${\mathtt T} =1.8$, $\omega = 1$ and $\epsilon = 0.025$.   \label{fig5}}	
\end{figure}

In order to see whether the map is positive or not, we look at the set of states to which the entire Bloch sphere of possible states of the detector are mapped to for different values of acceleration in Fig.~\ref{fig4}. We see that the dynamical maps is not positive and there are regions of the Bloch sphere that are mapped to nonphysical states under the action of the map. We also see that for large values of acceleration the size of this set of initial states that are mapped to nonphysical ones reduces, reflecting the behaviour of the second-smallest eigenvalue of $B_{0a}$ plotted in Fig.~\ref{fig:2b}. In a similar manner, the dependence of the nature of the map on the duration of inertial motion is shown in Fig.~\ref{fig5}. The dynamical map is not positive. We again see the correspondence with Fig.~\ref{fig:2a} with the size of set of states mapped to nonphysical ones increasing as the duration of inertial motion increases from $0.25$ to $2$.

\section{Conclusion}\label{conclusion}

We have studied the Unruh effect from the point of view of the open dynamics induced on the initially inertial and subsequently accelerated detector due to its interaction with the vacuum field. We see that the induced dynamics is not completely positive and in many cases it is not positive either. Lack of complete positivity is traced back to the correlations that build up between the detector and the field during its initial inertial motion. A similar scenario has been studied in terms of quantum master equations previously in~\cite{Sabrina2020} with focus on whether the induced dynamics is Markovian or not with the detector being always-on. When the duration of accelerated motion is finite, it was found that the dynamics is non-Markovian as observed by the increase in trace distance between pairs of states of the detector. The induced dynamics not being CP is also briefly discussed in~\cite{Sabrina2020} and our analysis clarifies its origin.  We observe here that the lack of complete positivity of the dynamics has to be taken into account when analysing the non-Markovian character of the open evolution since typical measures of non-Markovian behaviour assume that the open dynamics is CP~\cite{breuer_measure_2009,chruscinski_divisibility_2018}. 

Our results add one more consideration into the debate on whether NCP dynamical maps can be suitable mathematical descriptions of open quantum dynamics. The effect we are observing is purely kinematic in its origin and it involves only a detector that is switched on while it is at rest in an inertial frame and subsequently after a finite interval of time, subject to uniform acceleration. The trajectory of the detector is quite natural and reasonable with nothing contrived or forced about it. The two phases of its motion, inertial and accelerated provides a natural division demanding a mathematical description of the open dynamics during each phase separately. The description during the latter phase of motion, as we have seen, requires NCP dynamical maps.  

We have considered realistic detectors in terms them having a finite size. However, realistic considerations means that we have to also recognise that detection of the effects described, including signatures of NCP dynamics is going to be extremely challenging in a real experiment. As seen from Fig.~\ref{fig:1}, the difference between the eigenvalues of the dynamical maps with and without acceleration is very small. Even this small difference is obtained for accelerations of order 1, in units where $c=1$. These therefore represent substantial accelerations. Even the size we have considered for the detector of $\epsilon \sim 0.01$ is still quite large in the units we work with. On the other side of the coin, quantum systems typically do form exquisitely sensitive probes for any external forcing acting on them. Quantum metrology~\cite{barbieri_optical_2022,braunstein_statistical_1994,caves_quantum-circuit_2010,degen_quantum_2017,giovannetti_quantum_2006,helstrom_quantum_1976} is the science of using quantum resources like coherence and entanglement for pushing the limits of achievable precision in measurements. The detector we consider is a quantum system and it is quite conceivable that by using more than one detector in a joint quantum state, the precision required to detect the effects we have described may still be achievable in realistic experiments. Investigation of this question in detail is beyond the scope of the present work and is left as an open question to be addressed later.


\section*{Acknowledgments}
A.~S.~was supported by QuEST grant No Q-113 of the Department of Science and Technology, Government of India. The authors acknowledge the use of {\em Padmanabha} computational cluster of the centre for high performance computing at IISER Thiruvananthapuram. \\  

\appendix

\section{Wightman functions \label{AppA}}

Starting from the smeared field operator in \eqref{smearedfield} we expand the Wightman function as, 
\begin{equation}
   {\mathcal W}_M(\tau',\tau'') = \frac{1}{(2\pi)^{3}} \int \frac{d^{3}k}{2 \omega_{k}} g(\underline{k},\tau') g^{*}(\underline{k}, \tau'')
\end{equation}
where $g(\underline{k},\tau)$ is defined as in \eqref{gfunction}. Using this, we now calculate, 
\begin{eqnarray*}
      {\mathcal W}_M(\tau',\tau'') & = & \frac{1}{2(2\pi)^{3}} \int \frac{d^{3}k}{\omega_{k}} e^{i\underline{k} (\underline{x}_{\tau'} - \underline{x}_{\tau''}) + \epsilon \underline{k}(\underline{u}_{\tau'} + \underline{u}_{\tau''})}  \\
     & = & \frac{1}{2(2\pi)^{3}} \int \frac{d^{3}k}{\omega_{k}} e^{ik (x_{\tau'} - x_{\tau''} -  i \epsilon(\dot{x}_{\tau'} + \dot{x}_{\tau''}))} \\
         & &  \qquad \qquad \quad \times  e^{-i\omega_{k}(t_{\tau'} -t_{\tau''} -i\epsilon(\dot{t}_{\tau'} + \dot{t}_{\tau''}))}  \\
     & = & \frac{1}{4 \pi^2} \Big\{ -[t_{\tau'} - t_{\tau''} -i\epsilon(\dot{t}_{\tau'} + \dot{t}_{\tau''})]^2  \\
	& & \qquad  + [x_{\tau'} - x_{\tau''} -i \epsilon(\dot{x}_{\tau'} + \dot{x}_{\tau''})]^2 \Big\}^{-1}  \\
     & = &  \frac{1}{4 \pi^2}\Big\{{\big[ \underline{x}_{\tau^{\prime }} - \underline{x}_{\tau^{\prime \prime}} - i \epsilon (\underline{u}_{\tau^{\prime }} + \underline{u}_{\tau^{\prime \prime}})\big]^{2}} \Big\}^{-1}\\
\end{eqnarray*}
Where the integral over the $k$ vector is expanded as, 
\begin{equation}
    \label{k integral}
    \int d^{3}k \equiv \int_{0}^{2\pi} d\phi \int_{0}^{\pi} \sin{\theta} d\theta \int_{0}^{\infty}\omega_{k}^{2}d\omega_{k} 
\end{equation}
The Wightman function for inertial and uniformly accelerated motion can be derived by substituting the corresponding trajectories in \eqref{wightman2}. This has been discussed at length in \cite{Schlicht_2004}. Here instead, we calculate the new Wightman function ${\mathcal W}_M^{0a}$ in \eqref{cross_Wightman} that appear in integrals $\mathcal{Y}_{a0}^{\pm \pm}$ and $ \mathcal{Y}_{0a}^{\pm \pm}$ as a result of the combined inertial and accelerated trajectory we have chosen in this article. We can write the Wightman function as :
\begin{equation}\label{cross_appendix}
   {\mathcal W}_M^{0a}(\tau',\tau'') = \frac{1}{(2 \pi)^{3}} \int \frac{d^{3}k}{2\omega_{k}} g_{0}(\underline{k},\tau') g_{a}^{*}(\underline{k},\tau'')
\end{equation}
\\
Using \eqref{k integral}, the above integral simplifies as, 
\begin{eqnarray}
         {\mathcal W}_M^{0a}(\tau'\!,\!\tau'')\!& \!\! = \!\! & \frac{1}{8 \pi^{2}} \!\!\! \int_{0}^{\pi} \!\!\!\! d\theta\sin{\theta} \!\! \int_{0}^{\infty} \!\!\!\!\! d\omega_{k} \,\omega_{k} \,e^{i \omega_{k} [ t_{\tau''} - \tau' +i \epsilon(1 + \dot{t}_{\tau''}) ] } \nonumber \\
        & &  \qquad \qquad \qquad \times e^{- (x_{\tau''} + i \epsilon\dot{x}_{\tau''} ) \cos\theta  }    \nonumber  \\
        &\!\! = \!\! & -\frac{1}{8 \pi^{2}}\!\!\! \int_{0}^{\pi} d\theta \sin{\theta} \Big [ \big( t_{\tau''} - \tau' +i \epsilon(1 + \dot{t}_{\tau''}) \nonumber \\
        & & \qquad \qquad - (x_{\tau''} + i \epsilon\dot{x}_{\tau''} ) \cos\theta \big)^{2}  \Big ]^{-2} \nonumber \\
        & \!\! = \!\! & -\frac{1}{4\pi^{2}} \Big\{ \big[ \tau' - t_{\tau''} - i \epsilon( 1 + \dot{t}_{\tau''}) \big]^{2}\nonumber \\
        & &\qquad \qquad \qquad \quad- \big[ x_{\tau''} + i\epsilon \dot{x}_{\tau''}\big]^{2} \Big\}^{-1}
\end{eqnarray}
Where $x_{\tau''}$ amd $t_{\tau''}$ are defined by the accelerated trajectory in \eqref{accelcoord}. Similarly one can show, 
\begin{eqnarray}
    {\mathcal W}_M^{a0}(\tau',\tau'')&= & -\frac{1}{4\pi^{2}} \Big\{ \big[  t_{\tau'} -\tau'' - i \epsilon( 1 + \dot{t}_{\tau'}) \big]^{2} \nonumber \\
    & & \qquad \qquad \quad  - \big[ x_{\tau'} - i\epsilon \dot{x}_{\tau'} \big]^{2} \Big\}^{-1}
\end{eqnarray}

\bibliography{bib}
\bibliographystyle{ieeetr}

\end{document}